\pdfoutput=1
\documentclass[twocolumn,showkeys,showpacs,preprintnumbers,prd,superscriptaddress,nofootinbib]{revtex4-1}
\usepackage{graphicx}
\usepackage{epsf}
\usepackage{bm}
\usepackage{amsmath}
\usepackage{amsfonts}
\usepackage{amssymb}
\usepackage{epstopdf}
\usepackage{natbib}
\usepackage{hyperref}
\usepackage{color}
\usepackage{verbatim}
\usepackage{multirow}
\usepackage{bm}
\usepackage{hyperref}
\usepackage{url}
\usepackage{orcidlink}
\usepackage{wrapfig}
\usepackage{nicefrac, xfrac}
\usepackage{enumitem}
\usepackage[normalem]{ulem}
\usepackage{soul}

\definecolor{navyblue}{rgb}{0.0, 0.0, 0.5}
\definecolor{royalblue}{rgb}{0.25, 0.41, 0.88}
\definecolor{cadmiumgreen}{rgb}{0.0, 0.42, 0.24}
\definecolor{blue-violet}{rgb}{0.54, 0.17, 0.89}
\definecolor{darkviolet}{rgb}{0.58, 0.0, 0.83}
\definecolor{orange(colorwheel)}{rgb}{1.0, 0.5, 0.0}

\usepackage{hyperref}
\hypersetup{
    colorlinks=true, 
    linkcolor=royalblue, 
    citecolor=magenta}

\makeatletter\let\expandableinput\@@input\makeatother

\definecolor{WildStrawberry}{HTML}{EE2967}

\begin{document}

\title{Slow-rolling down the curvature: a reassessment of the Planck constraints on $\phi^2$ inflation in a closed universe}

\author{Enrico Specogna}
\email{especogna1@sheffield.ac.uk}
\affiliation{School of Mathematical and Physical Sciences, University of Sheffield, Hounsfield Road, Sheffield S3 7RH, United Kingdom} 

\author{Tatevik Vardanyan}
\email{tatevik@thp.uni-koeln.de}
\affiliation{University of Cologne, Faculty of Mathematics and Natural Sciences, Institute for Theoretical Physics, Cologne, Germany}

\author{William Giar\`e}
\email{w.giare@sheffield.ac.uk}
\affiliation{School of Mathematical and Physical Sciences, University of Sheffield, Hounsfield Road, Sheffield S3 7RH, United Kingdom} 

\author{Eleonora Di Valentino}
\email{e.divalentino@sheffield.ac.uk}
\affiliation{School of Mathematical and Physical Sciences, University of Sheffield, Hounsfield Road, Sheffield S3 7RH, United Kingdom} 

\begin{abstract}
\noindent We revisit the Cosmic Microwave Background (CMB) constraints on the spatial curvature of the Universe, assessing how they change when the curvature parameter and the primordial inflationary scalar spectrum are treated consistently within theoretically motivated frameworks. Instead of relying on the phenomenological parametrisation commonly used to capture curvature effects at the largest scales, we present a case study based on closed quadratic inflation, where the primordial spectrum is derived in full generality and in a gauge-invariant manner. Within this framework, we analyze both the \texttt{plik} PR3 and \texttt{CamSpec} PR4 Planck CMB likelihoods and find that the constraints on $\Omega_{\mathcal{K}}$ shift towards spatial flatness. In \texttt{plik} the preference for $\Omega_{\mathcal{K}}<0$ decreases from $\gtrsim 3.5\sigma$ to $\sim 2.5\sigma$, while in \texttt{CamSpec} it reduces further to $\sim 2\sigma$. At large angular scales ($\ell < 10$), our model explains the low-$\ell$ power suppression anomaly, notably improving the fit to the quadrupole. However, the reduced preference for highly negative values of $\Omega_{\mathcal{K}}$ only partially accounts for the lensing anomaly at high multipoles, worsening the fit to the \texttt{plik} spectrum at small scales. By contrast, in the \texttt{CamSpec} PR4 spectrum, where the lensing anomaly is less pronounced, the model yields an overall improvement. Our analysis highlights a key conceptual point: closed-inflation models tie the curvature parameter to the inflationary dynamics and the primordial spectrum, enforcing consistency conditions that do not necessarily allow for the large deviations from flatness seen in phenomenological parametrisations. In the case of quadratic inflation, these restrictions reduce the apparent evidence for negative curvature reported by earlier analyses, while allowing for a mildly closed geometry.
\end{abstract}



\maketitle


\section{Introduction}
\label{sec:introduction}

The spatially-flat $\Lambda$CDM model, built on the assumption of the validity of general relativity at cosmological scales and the existence of the two energy components contained in its name, $\Lambda$ (the cosmological constant) and CDM (cold dark matter), is to date our best attempt at unifying all of the cosmological observations currently available to us under the flag of a single paradigm. Despite its successes, this model suffers from some yet unresolved problems that put its validity into question -- see~\cite{CosmoVerse:2025txj} and the references therein for the most up-to-date summary of these issues. A notable incosistency is the \textit{Hubble tension}~\cite{Verde:2019ivm,DiValentino:2020zio,DiValentino:2021izs,Perivolaropoulos:2021jda,Schoneberg:2021qvd,Shah:2021onj,Abdalla:2022yfr,DiValentino:2022fjm,Kamionkowski:2022pkx,Giare:2023xoc,Hu:2023jqc,Verde:2023lmm,DiValentino:2024yew,Perivolaropoulos:2024yxv}, which refers to a non-trivial mismatch in the measurement of $H_0$, the present-day value of the Hubble parameter $H$, by different cosmological tracers. This tension is best exemplified by the astounding $6.2\sigma$ discrepancy between the measurement of $H_0$ obtained by the SH0ES team from Cepheid-calibrated type Ia supernovae~\cite{Riess:2021jrx,Breuval:2024lsv}, a late-time probe, and that from the cosmic microwave background (CMB) temperature and polarisation anisotropy spectra, an early-time probe, which the South Pole Telescope (SPT) ground-based observatory reports to be $H_0 = 66.66 \pm 0.60\, \rm km\,s^{-1}\,Mpc^{-1}$ in their recent  SPT-3G D1 release~\cite{SPT-3G:2025bzu}. If measurement systematics are to be ruled out as the root of the problems plaguing the $\Lambda$CDM scenario, such as the Hubble tension mentioned above~\cite{Efstathiou:2020wxn,Mortsell:2021nzg,Mortsell:2021tcx,Riess:2021jrx,Sharon:2023ioz,Murakami:2023xuy,Riess:2023bfx,Bhardwaj:2023mau,Brout:2023wol,Dwomoh:2023bro,Uddin:2023iob,Riess:2024ohe,Freedman:2024eph,Riess:2024vfa}, solutions need to be sought in alternative models of cosmology, by challenging the standard model's foundations.

An integral part of these foundations is the theory of \textit{inflation}~\cite{Guth:1980zm,Linde:1981mu}, commonly described by a minimally coupled scalar field, $\phi$, that slowly rolls down the slope of its potential, $\mathcal{V}(\phi)$. While providing a mechanism that explains the origin of the primordial density perturbations that went on to form the structures in the universe we now observe, inflation can also offer a solution to the so-called \textit{flatness problem}. First pointed out by Robert Dicke in 1969, the flatness problem is, in essence, a fine-tuning problem related to the amount of spatial curvature that characterises the universe: $\Omega_{\mathcal{K}}$, where the subscript $\mathcal{K}$ denotes the curvature parameter and can take the values $-1$, $0$, or $1$, depending on whether the universe is open, flat, or closed, respectively. The total energy budget of the universe can be defined as $\Omega = 8\pi G\rho / 3H^2$, with $\rho$ including the densities of matter, radiation, and $\Lambda$; for a flat universe, $\Omega = 1$. Dicke observed that since (1) $\Omega$ strays away from unity as the universe expands and (2) we currently observe the universe to be very close to flatness, or equivalently $\Omega_0 \approx 1$, then at early times the value of $\Omega$ must have been even closer to 1 than it is today -- an almost perfectly flat universe. To avoid such fine-tuning of the Universe’s initial conditions, we resort to inflation, which postulates an early period of exponential expansion able to guarantee a nearly flat Universe today for a wide range of initial values of $\Omega$.

State-of-the-art cosmological observations almost unanimously agree that the current universe, as far as we can tell, is spatially flat. The latest baryon acoustic oscillation (BAO) distance measurements from the Dark Energy Spectroscopic Instrument (DESI) yield $\Omega_{\mathcal{K}} = 0.025 \pm 0.041$~\cite{DESI:2025zgx}, in good agreement with previous SDSS BAO results, $\Omega_{\mathcal{K}} = 0.079^{+0.083}_{-0.10}$~\cite{eBOSS:2020yzd}. Both measurements are consistent with spatial flatness within $1\sigma$. The $3\times 2$pt measurements of the weak lensing of galaxies, their clustering, and the cross-correlation of the two can also be shown to agree with $\Omega_{\mathcal{K}} = 0$: the Dark Energy Survey (DES), when combined with CMB and Type Ia supernovae (SNIa) data, reports $\Omega_{\mathcal{K}} = 0.001 \pm 0.002$~\cite{DES:2022ccp}, while the KiDS-Legacy analysis finds $\Omega_{\mathcal{K}} = 0.011^{+0.054}_{-0.057}$~\cite{KiDS:2020ghu}. CMB data from the Atacama Cosmology Telescope (ACT) also indicate flatness to within $\sim 1.3\sigma$ (including Planck CMB and DESI BAO data)~\cite{ACT:2025tim}. Similarly, the CMB spectra of SPT alone constrain $\Omega_{\mathcal{K}}$ to be less than $1\sigma$ away from a null value~\cite{SPT-3G:2025bzu}. 

Perhaps the only discordant result, though entangled with many caveats and technicalities, comes from the temperature and polarisation spectra released by the Planck satellite, which still remains the most precise CMB probe. The Planck Release 3 (PR3), without including lensing reconstruction data, suggests evidence for a closed universe, with $\Omega_{\mathcal{K}} = -0.044^{+0.018}_{-0.015}$~\cite{Planck:2018vyg,DiValentino:2019qzk,Handley:2019tkm}, deviating from flatness at more than $3\sigma$. In the recent Planck Release 4 (PR4) this evidence is reduced, yet differences between likelihoods remain: \texttt{CamSpec}~\cite{Rosenberg:2022sdy} yields $\Omega_{\mathcal{K}} = -0.025^{+0.013}_{-0.010}$, still over $2\sigma$ away from $\Lambda$CDM, while \texttt{HiLLiPoP}~\cite{Tristram:2023haj} finds consistency with flatness to within $1\sigma$.

Inevitably, the presumed Planck preference for a curved Universe sparked (and continues to attract) a series of dedicated analyses, thereby~\cite{Park:2017xbl,Handley:2019tkm,Ryan:2019uor,DiValentino:2019qzk,Efstathiou:2020wem,DiValentino:2020hov,Benisty:2020otr,Vagnozzi:2020rcz,Vagnozzi:2020dfn,DiValentino:2020kpf,Yang:2021hxg,Bargiacchi:2021hdp,Cao:2021ldv,Dhawan:2021mel,Dinda:2021ffa,Zuckerman:2021kgm,Gonzalez:2021ojp,Akarsu:2021max,DiValentino:2022oon,DiValentino:2022rdg,Cao:2022ugh,Glanville:2022xes,Bel:2022iuf,Yang:2022kho,Stevens:2022evv,Vigneron:2022bgr,Jimenez:2022asc,Liu:2022mpj,Banerjee:2023rvg,Favale:2023lnp,Zhang:2023eup,Blachier:2023ooc,Qi:2023oxv,Giare:2023ejv,Foidl:2024xlv,Deng:2024uuz,Jensko:2024bee,Gariazzo:2024sil,Sanz-Wuhl:2024uvi,Amendola:2024gkz,Bhattacharya:2024hep,Shimon:2024mbm,Kuzmichev:2025fpm,Deng:2025uou,Wang:2025dzn}. This renewed interest in spatial curvature remains timely and relevant not primarily because of the aforementioned Planck preference for a closed Universe (an indication reduced in the PR4 release and, if taken at face value, not supported by most other early- and late-Universe probes), but because $\Omega_{\mathcal{K}}$ remains a parameter of central importance in many models extending beyond $\Lambda$CDM. These include, among others, proposed solutions to the Hubble tension~\cite{Sekiguchi:2020teg,Schoneberg:2021qvd,Schoneberg:2024ynd,Toda:2024ncp} and models of dynamical dark energy invoked to explain the present-day accelerated expansion of the Universe~\cite{Yang:2022kho,Bhattacharya:2024hep,Odintsov:2025sew,Hu:2024niv}.

In almost all studies characterizing spatial curvature, including the Planck analyses themselves, the primordial spectrum $\mathcal{P}_{\zeta}$ in a non-flat universe is modeled using the following heuristic formula~\cite{Lyth:1990dh,Ratra:1994vw,Efstathiou:2003hk,Park:2017xbl,Efstathiou:2020wem,Asorey:2025hgx}:
\begin{equation}
\label{eq:spec_efs}
    \mathcal{P}_{\zeta} \propto \frac{(n^2 - 4\mathcal{K})^2}{n(n^2 - \mathcal{K})^2} A_s \left(\frac{k}{k_*}\right)^{n_s - 1},
\end{equation}
where 
\begin{equation}
    n^2 = k^2 + \mathcal{K}
    \label{eq:n_vs_k}
\end{equation}
takes discrete values such that $n \geq 3$ when $\mathcal{K} = 1$. The pivot scale is usually fixed at $k_* = 0.05\,\mathrm{Mpc}^{-1}$, which corresponds to the scale at which the amplitude $A_s$ and the tilt $n_s$ of the primordial spectrum are evaluated. 

Eq.~(\ref{eq:spec_efs}) is an adaptation of the power-law expression generally used to test flat models, which can be recovered by fixing $\mathcal{K} = 0$. The $n$-dependent prefactor is introduced to heuristically capture the modifications induced by curvature on the largest scales, where a closed geometry leads to discrete eigenmodes and non-trivial deviations from the flat case. While this construction is convenient, it lacks a direct derivation from a concrete inflationary model and should be interpreted with caution, as depending on the underlying model it may not adequately reproduce the imprints of the inflationary dynamics on the primordial spectrum at those scales. This limitation has been discussed in several works that attempted to move beyond such a heuristic treatment~\cite{Ratra:2017ezv,Handley:2019anl,Zhang:2003eh,Thavanesan:2020lov,Vigneron:2024bfj}.\footnote{For discussions surrounding inflation in a closed Universe, see also Refs.~\cite{Abbott:1986,Gerlach:1978,Wolf:1988mw,Wolf:1990kj,PoncedeLeon:1990gg,Schmidt:1993dq,Linde:1995xm,Pavluchenko:2001mt,Ellis:2001ym,Linde:2003hc,Uzan:2003nk,Pavluchenko:2003ge,Lasenby:2003ur,Green:2007gs,Cicoli:2010ha,Kleban:2012ph,Ratra:2017ezv,Ooba:2017ukj,Gordon:2020gel,Murata:2021vnb,Motaharfar:2021gwi,Motaharfar:2021gwi,Bonga:2016cje,Bonga:2016iuf,DAgostino:2023tgm,Muller:2023tsh}.}

Yet another source of concern, besides the heuristic nature of Eq.(\ref{eq:spec_efs}), is that, as emphasised in Ref.~\cite{PeterUzan2009} (see also~\cite{Mukhanov:1990me,Abbott:1986,Gerlach:1978}), predictions for a closed universe can vary significantly across different gauges, especially on large scales corresponding to wavelengths comparable to the curvature radius. Broadly speaking, this is due to the fact that, in a closed FRW geometry ($\mathcal K>0$), scalar perturbations are expanded in hyperspherical harmonics on $S^3$ with discrete wavenumbers $n$. The physically propagating modes start at $n \geq 3$, while the $n=1,2$ sectors are non-dynamical or pure gauge. At the largest scales, corresponding to small $n$, the usual time slicings (e.g., Newtonian, synchronous, comoving) differ markedly, and the constraint equations introduce $\mathcal O(\mathcal K/n^2)$ terms that mix scalar-field and metric perturbations. As a consequence, spectra computed for gauge-fixed variables (such as $\delta\phi$ in comoving gauge or the Bardeen potentials in Newtonian gauge) can exhibit gauge-dependent amplitudes and tilts precisely where curvature effects are strongest.\footnote{This issue is further compounded by the choice of initial vacuum in a closed geometry, where there is no exact Minkowski sub-horizon limit and naive Bunch–Davies prescriptions can pick up gauge-specific normalisations on large scales.} Therefore, alternative formulations and analyses performed within a fixed gauge -- even when aiming to go beyond this parametrisation and provide more precise calculations -- may still carry residual ambiguities rooted in the gauge choice.

Although these issues may seem like technical subtleties of mainly academic interest, they can become highly relevant when deriving observational constraints on the spatial curvature parameter from Planck data. The apparent Planck preference for a closed universe arises not only from the well-known lensing anomaly\footnote{The lensing anomaly is usually quantified via the phenomenological parameter $A_L$, which rescales the spectrum of the lensing potential affecting the primary CMB anisotropies~\cite{Calabrese:2008rt}. In the Planck PR3 spectra, $A_L$ is measured to be $1.180 \pm 0.065$~\cite{Planck:2018vyg}, roughly $3\sigma$ away from its $\Lambda$CDM value of $A_L = 1$. This systematic effect appears to drive the preference of Planck data not only for a closed universe~\cite{DiValentino:2019qzk,Handley:2019tkm,Vagnozzi:2020rcz,DiValentino:2020hov}, but also for modified gravity scenarios~\cite{DiValentino:2015bja,Specogna:2023nkq,Specogna:2024euz,Giare:2025ath}.} -- i.e. the preference for enhanced smoothing of the acoustic peaks at small scales~\cite{DiValentino:2015bja,Renzi:2017cbg,Domenech:2020qay,Addison:2023fqc,Kable:2020hcw,Addison:2015wyg} -- but also from small yet important differences in the fit to the CMB power spectra at the largest scales~\cite{Efstathiou:2003hk,Bonga:2016iuf}. Planck measurements of temperature and polarisation anisotropies indeed reveal a series of long-standing anomalies in this regime: the TT spectrum at the lowest multipoles $\ell \lesssim 10$ (notably the quadrupole and octopole) lies systematically below the best-fit $\Lambda$CDM prediction~\cite{Planck:2018vyg,Planck:2019nip}. It is precisely on these scales that the heuristic $n$-dependent corrections in Eq.~(\ref{eq:spec_efs}) -- and, more broadly, gauge-dependent treatments in alternative formulations -- can mimic a preference for negative curvature, thereby potentially contributing to the apparent Planck signal for $\Omega_{\mathcal{K}} < 0$ and raising doubts about the robustness of curvature constraints derived from gauge-dependent approaches.

In this work, we reassess the implications of the above limitations by adopting the primordial power spectrum for a closed universe originally derived in Ref.~\cite{Kiefer:2021iko}. Specifically, we employ a gauge-invariant variable, constructed in analogy with the Mukhanov–Sasaki variable, which combines scalar field and metric perturbations to yield a fully gauge-invariant expression for the primordial spectrum in a closed geometry. We use this theoretically motivated spectrum as a case study to evaluate the impact on current constraints on $\Omega_{\mathcal{K}}$ from Planck data and to reassess the presumed preference for a closed universe within physically motivated models of closed inflation. The details of the derivation are summarised in Sec.~\ref{sec:theory}, the data used in our analysis are described in Sec.~\ref{sec:data}, the results are presented in Sec.~\ref{sec:results}, and our conclusions on the flatness of the present universe are drawn in Sec.~\ref{sec:conclusion}.

\section{The Gauge-invariant Primordial power Spectrum in a Closed Universe}
\label{sec:theory}

In a closed Friedmann–Lemaître–Robertson–Walker (FLRW) universe, we consider an inflationary phase driven by a scalar field $\phi$ with a quadratic potential $\mathcal{V}(\phi) = \frac{1}{2} m^2 \phi^2$. As one of the simplest and most widely used slow-roll potentials, it provides a standard benchmark for studying inflationary dynamics and for isolating the effects induced purely by the spatial curvature in a closed FLRW background.

The line element of the FLRW metric in conformal time $\eta$ is given by
\begin{equation}
\begin{aligned}
ds^2 = a^2(\eta)\left[-d\eta^2 + d\chi^2 + s^2_{\mathcal{K}}(\chi)\, d\Omega^2 \right],
\end{aligned}
\end{equation}
where $a(\eta)$ is the scale factor, $\chi$ is the comoving radial distance, and $d\Omega^2 = d\theta^2 + \sin^2{\theta}\, d\phi^2$ is the infinitesimal solid angle. For the closed model, the function $s_{\mathcal{K}}(\chi)$ takes the form
\begin{equation}
s_{\mathcal{K}}(\chi) = \frac{\sin\left(\sqrt{|\mathcal{K}|}\, \chi\right)}{\sqrt{|\mathcal{K}|}},
\end{equation}
with curvature parameter $\mathcal{K} = +1$.

In the slow-roll regime, the scalar field satisfies the following conditions:
\begin{equation}
    \dot{\phi}^2 \ll \mathcal{V}(\phi), \quad |\ddot{\phi}| \ll 3 H |\dot{\phi}|.
\end{equation}
The slow-roll parameters are defined as
\begin{equation}
    \varepsilon = 1 - \frac{\mathcal{H}'}{\mathcal{H}^2}, \quad
    \delta = \varepsilon - \frac{\varepsilon'}{2\mathcal{H}\varepsilon},
\end{equation}
where $\mathcal{H}$ denotes the comoving Hubble parameter, related to the Hubble parameter by $\mathcal{H} = aH$. The prime indicates a derivative with respect to conformal time $\eta$.

The power spectrum in the slow-roll approximation using a gauge-invariant variable, originally derived in Ref.~\cite{Kiefer:2021iko}, is summarised here together with the assumptions underlying its derivation. The calculation proceeds within the framework of canonical quantum gravity. Implementing canonical quantization for the perturbed inflationary universe model leads to the corresponding Wheeler–DeWitt equation. Applying a semiclassical Born–Oppenheimer type of approximation to this equation allows one to recover Schr\"{o}dinger equations for the perturbation modes (see Eq.~(57) in Ref.~\cite{Kiefer:2021iko}), together with the quantum-gravitationally corrected Schr\"{o}dinger equation at the next order of approximation. From the resulting Schr\"{o}dinger equation, the power spectrum for the gauge-invariant generalised Mukhanov-Sasaki variable is obtained (see Eq.~(74) in Ref.~\cite{Kiefer:2021iko}). The initial state of the perturbation modes is taken to be the adiabatic vacuum, the standard choice for quantum fluctuations in this framework. The quantum-gravitationally corrected Schr\"{o}dinger equation and the corresponding corrected spectrum are derived in Ref.~\cite{Vardanyan:2023ldj}.

\begin{figure*}[htpb!]
\centering
\includegraphics[width=\textwidth]{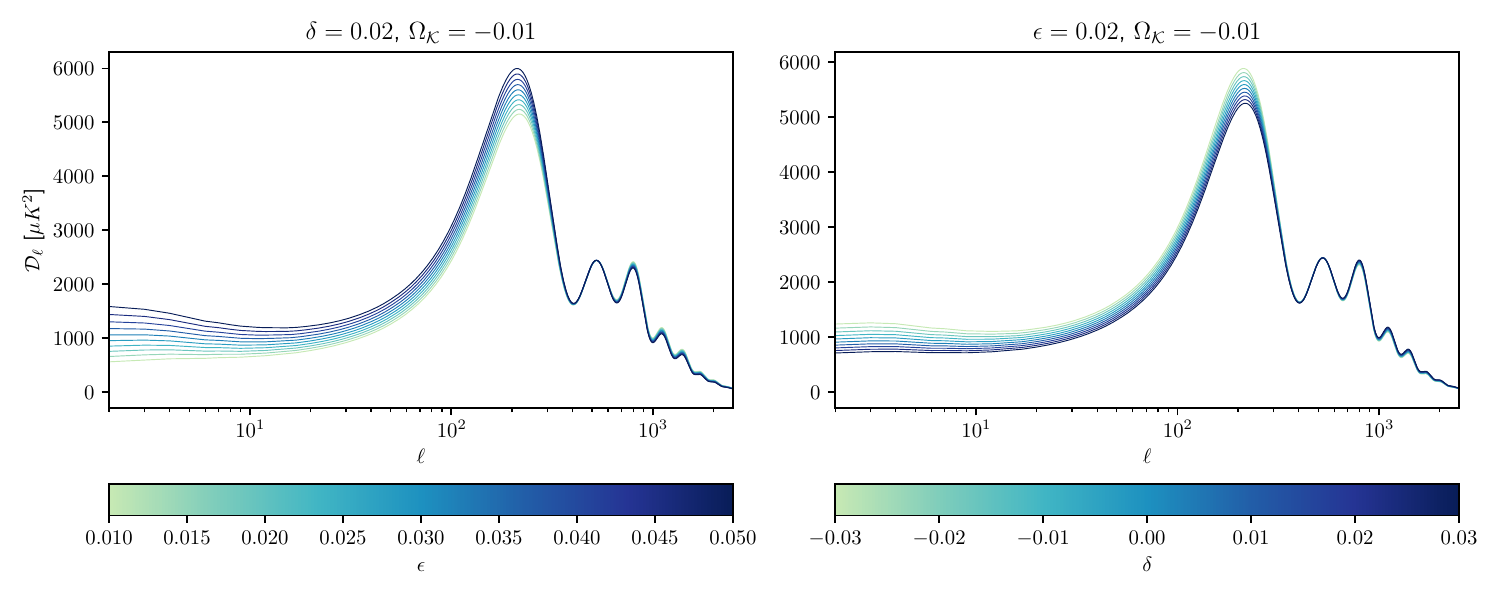}
\caption{Left panel: Variations of $C_\ell^{TT}$ when the slow-roll parameter $\epsilon$ is varied in the range $\epsilon = [0.01, 0.05]$ with fixed $\delta = 0.02$ and $\Omega_K = -0.01$. Right panel: Variations of $C_\ell^{TT}$ when the slow-roll parameter $\delta$ is varied in the range $\delta = [-0.03, 0.03]$ with fixed $\epsilon = 0.02$ and $\Omega_K = -0.01$.}\label{fig:eps-delta}
\end{figure*}

In addition to the power spectrum of the generalised Mukhanov--Sasaki variable, the power spectrum of the gauge-invariant variable $\zeta_{\rm BST}$ (named after Bardeen, Steinhardt, and Turner~\cite{bardeen1983}) is also obtained in Ref.~\cite{Kiefer:2021iko} (see Eq.~(79) therein), where it is expressed in terms of the former. The variable $\zeta_{\rm BST}$ remains conserved during the evolution of super-Hubble modes in non-flat FLRW models~\cite{PeterUzan2009}; hence, it is the natural quantity for relating the primordial scalar perturbations at the end of inflation to the temperature anisotropies of the CMB and the quantity that directly enters as input in our Boltzmann code computations.\\

In obtaining the power spectrum, the derivation was performed to first order in the slow-roll parameters $\varepsilon$ and $\delta$, with higher-order terms neglected. In addition, curvature corrections involving higher powers of $\mathcal{K}/\mathcal{H}^2$ were omitted, a justified approximation since during inflation the comoving Hubble horizon $\mathcal{H}^{-1}$ decreases rapidly.

The primordial power spectrum for $\zeta_{\rm BST}$ takes the following explicit form\footnote{Here, the Planck mass is defined as $M_{\rm P} := \sqrt{1/G}$, and calculations are performed in natural units, $\hbar = c = 1$.},\footnote{
For convenience and consistency with our numerical implementation, we write the spectrum in the compact, power-law-like form commonly used in standard slow-roll treatments (e.g. Eq.~(8.226) of~\cite{PeterUzan2009}). This expression is algebraically equivalent to the logarithmic form given in Eqs.~(74) and (77) of Ref.~\cite{Kiefer:2021iko} to first order in the slow-roll parameters which is the order to which the original derivation applies.
}:
\begin{equation}
\label{eq:spec_full}
\begin{aligned}
 \mathcal{P}_{\zeta_{\rm BST}} = \ & \frac{1}{\pi} \frac{H^2}{M_P^2\left(\varepsilon + \frac{\mathcal{K}}{\mathcal{H}^2}\right)} \frac{1 - \mathcal{K}/n^2}{f_\varepsilon^3(n)} \Bigg[1 + \frac{\mathcal{K} \left(\varepsilon + \mathcal{K}/\mathcal{H}^2\right)}{n^2 - 4\mathcal{K}} \Bigg] \\
 & \times \Tilde{C}_{\varepsilon, \delta} \left( \frac{n}{a H} f_\varepsilon(n)\right)^{-2(\gamma - 3\lambda_{\varepsilon,\delta}(n)/2)},
\end{aligned}    
\end{equation}
with
\begin{equation}
\tilde{C}_{\varepsilon,\delta} = 1 - 2\varepsilon + \left( \gamma - \frac{3}{2} \lambda_{\varepsilon,\delta}(n) \right)(4 - 2\gamma_E - 2\ln(2)),
\label{Ctilde}
\end{equation}
where $\gamma = 2\varepsilon - \delta$, and $\gamma_E \simeq 0.5772$ is the Euler--Mascheroni constant. 
The functions $f_\varepsilon(n)$ and $\lambda_{\varepsilon,\delta}(n)$ are given by
\begin{equation}
\begin{aligned}
  f_\varepsilon(n) = \Bigg\{ 1 - \frac{\mathcal{K}}{n^2} 
  & - \frac{\mathcal{K}}{n^2 - 4 \mathcal{K}} \Bigg( 2 + 3\varepsilon + \frac{(3\varepsilon - 1) \mathcal{K}}{n^2} \Bigg) \\
  & + \frac{3(3 - 2\varepsilon) \mathcal{K}^2}{\left(n^2 - 4\mathcal{K}\right)^2} \left( 1 - \frac{\mathcal{K}}{n^2} \right) \Bigg\}^{\frac{1}{2}},
\end{aligned}
\label{f1}
\end{equation}
and
\begin{equation}
\begin{aligned}
 \lambda_{\varepsilon,\delta}(n)=1-\Bigg\{&1-\frac{8\mathcal{K}}{n^2-4 \mathcal{K}}\frac{1}{4\gamma+3}\\
  &\times \Bigg(3\gamma+12\delta+\frac{3\varepsilon}{2}\frac{n^2-\mathcal{K}}{n^2-4\mathcal{K}}\Bigg)\Bigg\}^{\frac{1}{2}}.
\end{aligned}
\end{equation}

These functions arise when solving the Schr\"{o}dinger equation for the perturbation modes in a closed FLRW background: $f_\varepsilon(n)$ appears in the argument of the Bessel functions, and $\lambda_{\varepsilon,\delta}$ determines their order, encoding the corrections induced by spatial curvature in a closed FLRW geometry (see Eq.~(63) in Ref.~\cite{Kiefer:2021iko}).

Furthermore, let us recall the power spectrum in the flat case~\cite{Brizuela:2016gnz}, which can be recovered from the power spectrum in Eq.~\eqref{eq:spec_full} by taking the large-wavenumber limit:
\begin{align}
\mathcal{P}^{\rm f}_{\zeta}(k) = \frac{1}{\pi} \frac{H^2}{M_P^2 \varepsilon} C_{\varepsilon,\delta} \left( \frac{k}{aH} \right)^{-2\gamma},
\label{flatp}
\end{align} 
where
\begin{equation}
C_{\varepsilon,\delta} = 1 - 2\varepsilon + \gamma(4 - 2\gamma_E - 2\ln(2)).
\end{equation} 

Here, the power spectrum is expressed in terms of the conserved gauge-invariant variable $\zeta$ for the flat FLRW model. Its relation to $\zeta_{\rm BST}$ is discussed in~\cite{PeterUzan2009}.\\

We stress that this derivation of the primordial spectrum based on Ref.~\cite{Kiefer:2021iko}, is carried out for a quadratic inflationary background. The background equations are first solved explicitly for the potential $V(\phi)=m^{2}\phi^{2}/2$, and the resulting functions $H(t)$, $a(t)$, and $\phi(t)$ are then inserted into the perturbation equations. Therefore, although the final expressions are written in terms of slow-roll parameters, the resulting spectrum is not universal in a closed Universe: it is valid only for the quadratic potential used in the derivation. This model dependence is not merely a byproduct of methodology, but reflects an intrinsic obstruction specific to the closed case that has no analogue in flat  inflation. When $\mathcal{K}>0$, the Friedmann equation contains the additional contribution $\mathcal{K}/a^{2}$, whose evolution cannot be rewritten in terms of $\epsilon$, $\delta$, and their time derivatives. Unlike the slow-roll terms, the curvature contribution does not scale with the slow-roll hierarchy and depends explicitly on the full background solution through $a(t)$. The suppression of  $\mathcal{K}/a^{2}$, the duration of the curvature-dominated regime, and the onset of the slow-roll phase all depend sensitively on the detailed form of $V(\phi)$. As a result, two different potentials with identical values of $(\epsilon,\delta)$ can lead to different background evolutions in the closed case. This already prevents the Mukhanov--Sasaki frequency from being written solely in terms of slow-roll parameters, in contrast with the flat scenario. In addition, the curvature-modified Mukhanov--Sasaki equation contains contributions proportional to $\mathcal{K}$ and to the discrete Laplacian eigenvalues on $S^{3}$, $n(n+2)$. These terms involve $a$, $a'$, and $a''$ in ways that cannot be encoded by the slow-roll hierarchy alone. This is a purely geometric effect of the closed spatial topology: the discrete spectrum of scalar harmonics and the curvature-dependent contributions couple directly to the background evolution, introducing an explicit dependence on $V(\phi)$.\\

We note that~\cite{Kiefer:2021iko} demonstrated that the power spectrum in a closed FLRW model is suppressed at large scales and exhibits a cutoff at the largest scales, compared to the flat case, due to the finiteness of the spatial geometry. It should also be pointed out that the power spectrum in Eq.~\eqref{eq:spec_full} exhibits an explicit and prominent scale dependence at large scales due to the curvature of the universe. Consequently, the usual power-law approximation, which assumes only weak scale dependence quantified by the spectral index, is not applicable in this case.
 
\section{Methodology and datasets}
\label{sec:data}

In this section, we describe the numerical implementation of the inflationary model (Sec.~\ref{sec:data.parameterization}) and discuss how we test its viability against CMB data (Sec.~\ref{sec:data.data}).

\subsection{Re-expressing $\mathcal{P}^c_\zeta$} \label{sec:param}
\label{sec:data.parameterization}

We first expand on how Eq.~\eqref{eq:spec_full} was implemented into the Boltzmann solver \texttt{CAMB}~\cite{Lewis:1999bs, Howlett:2012mh} for the calculation of the observables used in this analysis, namely the temperature, polarisation, and CMB lensing potential spectra.

In the flat $\Lambda$CDM model, the mild scale dependence of the primordial power spectrum can be rewritten in the form of a power law:
\begin{equation}
\label{eq:power_law}
    \mathcal{P}^f_\zeta(k) = A_s \left( \frac{k}{k_*} \right)^{n_s - 1},
\end{equation} 
which can be used as a parametric approximation of Eq.~\eqref{flatp}.  Because Eq.~\eqref{eq:power_law} is the mathematical form employed by \texttt{CAMB} to evaluate $\mathcal{P}^f_\zeta$, it would be natural to also re-express Eq.~\eqref{eq:spec_full} in terms of a power law, by calculating what $A_s$ and $n_s$ would look like in our model. However, as explained in Sec.~\ref{sec:theory}, this is not trivial due to the strong $k$-dependence of the primordial spectrum in this model.  We therefore adopt the following strategy to implement Eq.~\eqref{eq:spec_full}. 

\begin{figure}[htpb!]
\centering
\includegraphics[width=\columnwidth]{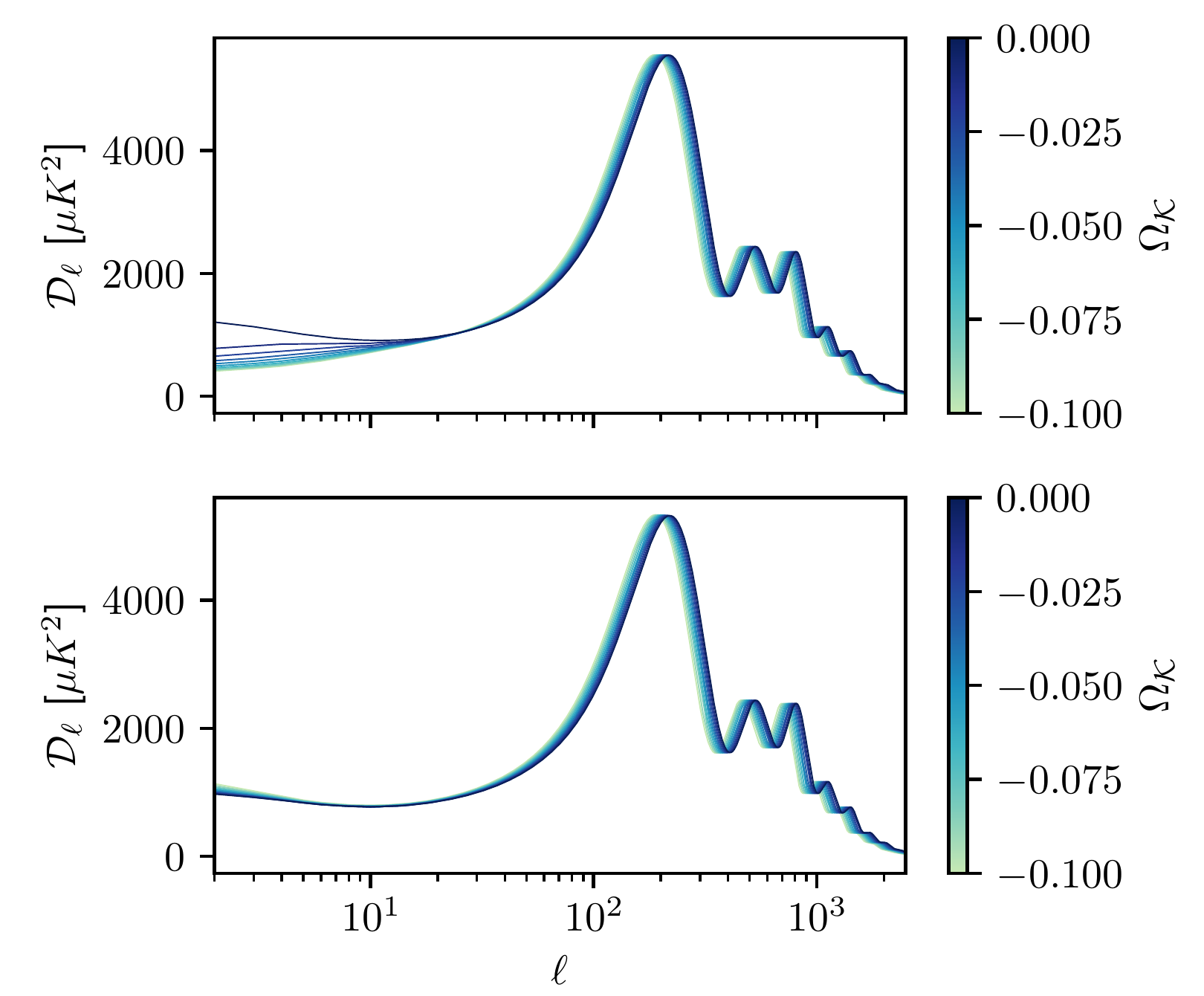}
\caption{The CMB TT spectrum according to Eq.(\ref{eq:spec_param}) (upper figure) and the one calculated with the standard \texttt{CAMB} implementation (lower figure). The lower $\Omega_{\mathcal{K}}$, the stronger the suppression of the low-$\ell$ spectrum.}\label{fig:comp-omegak}
\end{figure}

First, due to the complexity of the full expression in Eq.~\eqref{eq:spec_full}, we introduce the following approximations to simplify its implementation in \texttt{CAMB}~\cite{Lewis:1999bs, Howlett:2012mh}: we set the slow-roll parameters to zero within the functions $f_\varepsilon(n)$ and $ \lambda_{\varepsilon,\delta}(n)$, where their contributions are verified to be negligible. Thus, for convenience, in the present analysis, we consider the power spectrum in Eq.~\eqref{eq:spec_full} with
\begin{equation}
\label{eq:simply}
    f(n):=\lim_{\varepsilon\rightarrow 0}f_\varepsilon(n) \quad \textnormal{and} \quad \lim_{\varepsilon,\delta\rightarrow 0} \lambda_{\varepsilon,\delta}(n) = 0.
\end{equation}
The function in Eq.~\eqref{Ctilde} also simplifies to $\tilde{C}_{\varepsilon,\delta}(n)\approx C_{\varepsilon,\delta}$. We denote the resulting expression for the power spectrum in the closed FLRW universe by $\mathcal{P}^{\rm c}_{\zeta}(n)$:
\begin{equation}
\begin{aligned}
  \mathcal{P}^c_{\zeta}(n)=\ & \frac{1}{\pi}\frac{H^2 \left(1-\mathcal{K}/n^2\right)}{M_P^2\left(\varepsilon+\frac{\mathcal{K}}{\mathcal{H}^2}\right)f^{3}(n)}\Bigg[1+\frac{\mathcal{K}}{n^2-4\mathcal{K}}\left(\varepsilon+\frac{\mathcal{K}}{\mathcal{H}^2}\right)\Bigg]\\
  & \times C_{\varepsilon,\delta} \left(\frac{n}{a H}f(n)\right)^{-2\gamma}. 
   \label{fsp1}
\end{aligned} 
\end{equation}

The approximation introduced by Eq.~(\ref{eq:simply}) is not based on a strict order-by-order expansion of the parameters $\epsilon$ and $\mathcal{K}/n^2$ (the combination of which is neglected in $f_{\epsilon}(n)$ and $\lambda_{\epsilon,\delta}(n)$ but still appears in Eq.~(\ref{fsp1})), but rather on numerical relevance for the CMB observables. This simplification was designed so that the modified $\mathcal{P}^c_{\zeta}(n)$ can be rewritten as a multiplicative correction to the standard power-law spectrum as explained in the following lines, avoiding numerical instabilities in \texttt{CAMB}. This choice aims to increase the speed of convergence of the MCMC scans while ensuring a negligible impact on the resulting cosmological constraints.\\

Second, we generalize the primordial spectrum to include our non-flat scenario by introducing a factor, $\Phi(n, \epsilon, \delta, \Omega_k)$, that reduces to unity when $\Omega_k = 0$:
\begin{equation}
\label{eq:spec_param}
    \mathcal{P}^{\rm c}_{\zeta}(n) = \mathcal{P}^{\rm f}_{\zeta}(n)\, \Phi(n, \epsilon, \delta, \Omega_k),
\end{equation}
where $\mathcal{P}^f_{\zeta}(n)$ simply results from redefining the wavenumber in Eq.~\eqref{eq:power_law} according to Eq.~\eqref{eq:n_vs_k}. 
Here, $\Phi(n, \epsilon, \delta, \Omega_k)$ is not just an arbitrary function that needs to be constrained or reconstructed by the data.
As a matter of fact, it is precisely defined as the ratio between Eq.~\eqref{fsp1} and its equivalent expression in the $\Lambda$CDM limit, Eq.~\eqref{flatp}:
\begin{equation}
\begin{aligned}
 \Phi(n, \epsilon, \delta, \Omega_k) =\ &\frac{\mathcal{P}^{\rm c}_{\zeta}(n)}{\mathcal{P}^{\rm f}_{\zeta}(n)} \\
 =\ &\frac{\varepsilon}{\left(\varepsilon + \frac{\mathcal{K}}{\mathcal{H}^2}\right)}\left(1 - \frac{\mathcal{K}}{n^2}\right) \\
 &\times \Bigg[1 + \frac{\mathcal{K}}{n^2 - 4\mathcal{K}}\left(\varepsilon + \frac{\mathcal{K}}{\mathcal{H}^2}\right)\Bigg] f(n)^{-2\gamma - 3},
\end{aligned}
\end{equation}
where the dependence on $\Omega_k$ enters through the dimensionless curvature combination $\mathcal{K}/\mathcal{H}^2$. By construction, $\Phi(n,\epsilon,\delta,\Omega_k)$ reduces to unity in the flat limit $\mathcal{K}/\mathcal{H}^2\to0$, recovering the flat-universe case. Conversely, in a closed universe ($\mathcal{K}=1$), the flat contribution cancels out and the expression isolates the purely curvature-dependent corrections. In this way, the result can be interpreted as a ``correction factor’’ to the familiar power law of Eq.~\eqref{eq:power_law}, which makes its implementation into \texttt{CAMB} straightforward.\\

In order to check whether the approximations introduced by Eq.~(\ref{eq:simply}) are robust and can be safely used to constrain $\mathcal{P}_{\zeta}$ with data, in Appendix~\ref{appendix:B} we carry out a comparison between the approximated (i.e., Eq.~(\ref{fsp1})) and the full (i.e., Eq.~(\ref{eq:spec_full})) versions of $\mathcal{P}_{\zeta}$ by implementing both in \texttt{CAMB}. The difference between the approximated and non-approximated spectra of the CMB temperature anisotropies is shown to be well within the uncertainties of Planck's 2018 measurements of the temperature spectrum~\cite{Planck:2018vyg}, meaning that choosing Eq.~(\ref{fsp1}) over Eq.~(\ref{eq:spec_full}) does not introduce any relevant systematic shift in the constraints for our model outlined in Sec.~\ref{sec:results}.\\

Finally, the effect of varying the slow-roll parameters $\epsilon$ and $\delta$ in this setup is illustrated in Fig.~\ref{fig:eps-delta}. Increasing $\epsilon$ leads to an overall enhancement of the CMB temperature power spectrum, particularly at low multipoles, while increasing $\delta$ produces the opposite effect, suppressing power across all scales. In Fig.~\ref{fig:comp-omegak}, instead, we compare the TT spectrum obtained from our model using Eq.~\eqref{eq:spec_param} to the standard \texttt{CAMB} implementation of curvature. While both approaches yield similar results at high multipoles (the damping tail), the suppression of low-$\ell$ power, especially the quadrupole, is significantly stronger in our model. The standard implementation shows only a mild suppression at low-$\ell$ for the same values of $\Omega_{\mathcal{K}}$.

\subsection{Planck Likelihoods and Data}
\label{sec:data.data}

The $\Lambda$CDM parameter space -- hereafter referred to as $\lambda$ -- is expanded by the additional parameters introduced by our model: $\lambda \longrightarrow \lambda_c = \lambda \cup \{\epsilon, \delta, \Omega_k\}$.\footnote{$\lambda$ includes the energy densities of baryonic ($\Omega_{\rm b} h^2$) and dark ($\Omega_{\rm c} h^2$) matter, the optical depth integrated to the reionization epoch ($\tau$), the sound horizon-angular diameter distance ratio calculated at last scattering ($\theta_{\rm MC}$) as well as the amplitude ($A_s$) and tilt ($n_s$) of the flat primordial scalar spectrum~\eqref{eq:power_law} employed in Eq.~\eqref{eq:spec_param}. $n_s$ is, in this case, a derived parameter, as it depends on $\epsilon$ and $\delta$: $n_s=2\delta-4\epsilon$.}

To efficiently sample $\lambda_c$ within the prior volume given by the flat priors listed in Tab.~\ref{tab:priors}\footnote{Note that throughout this work we assume a prior $\epsilon>0$. In canonical single-field 
inflation this follows from the null-energy condition, which fixes $\dot H$  to be negative in flat space. In closed geometries the curvature term can slightly modify its magnitude, but it cannot flip its sign without breaking the slow-roll dynamics: if curvature dominated $\dot H$, the evolution would 
no longer be controlled by the inflaton field and the system would not describe a genuine slow-roll phase. Such configurations therefore lie outside the class of models considered here.}, we use the \texttt{Cobaya} Markov Chain Monte Carlo (MCMC) package~\cite{Torrado:2020dgo}. The sample chains obtained in this fashion are then analyzed using the \texttt{Getdist} plotting code~\cite{Lewis:2019xzd}, so as to derive the constraints on $\lambda_c$ and their graphical 1D and 2D representations, presented in Sec.~\ref{sec:results}.

The viability of the inflationary model outlined in Sec.~\ref{sec:theory} is tested  using exclusively CMB data: both the curvature and the slow-roll parameters $\epsilon$ and $\delta$ directly affect the primordial spectrum $\mathcal{P}_{\zeta}$, which in turn determines the temperature anisotropies of the CMB, as illustrated in Fig.~\ref{fig:eps-delta} and Fig.~\ref{fig:comp-omegak}. This makes the Planck CMB measurements a natural probe to constrain the model. The Planck CMB likelihoods employed in our analysis, along with their corresponding keys, are listed below.

\begin{table}[htpb!]
    \centering
    \renewcommand{\arraystretch}{1.2}
    \resizebox{0.766 \columnwidth}{!}{%
    \begin{tabular}{l @{\hspace{2cm}} c}
        \hline\hline
        Parameter & Prior \\
        \hline
        $\Omega_{\rm b} h^2$       & $[0.017,\ 0.027]$ \\
        $\Omega_{\rm c} h^2$       & $[0.09,\ 0.15]$ \\
        $\tau$                     & $[0.01,\ 0.2]$ \\
        $\log(10^{10} A_s)$        & $[2.6,\ 3.5]$ \\
        $100\,\theta_{\rm MC}$     & $[1.03,\ 1.05]$ \\
        $\Omega_{\mathcal{K}}$          & $[-0.05,\ 0]$ \\
        $\epsilon$                 & $[0,\ 0.05]$ \\
        $\delta$                   & $[-0.03,\ 0.03]$ \\
    \hline\hline
    \end{tabular}
    }
\caption{Flat priors adopted to constrain the parameters of the inflationary model presented here, including $\epsilon, \delta, \Omega_k$.}
\label{tab:priors}
\end{table}

\begin{itemize}
    \item \textbf{PL18}: this is a likelihood combination including \texttt{plik}~\cite{Planck:2019nip} for the temperature (TT), polarization (EE), and temperature-polarization cross-correlation (TE) Planck 2018 spectra at $\ell > 30$, while in the $\ell < 30$ range we use the \texttt{simall} likelihood~\cite{Planck:2018vyg} for E-mode polarization measurements, and the \texttt{Commander} likelihood~\cite{Planck:2018yye} for the TT anisotropies.
    \item \textbf{lensing}: this is the lensing spectrum, $C_{\ell}^{\phi\phi}$, from the 2018 data release of Planck's maps~\cite{Planck:2018lbu}; $\phi$ is the lensing potential that can be reconstructed through quadratic estimators of TT and EE.
    \item \textbf{CamSpec}: this combination includes the \texttt{CamSpec}~\cite{Rosenberg:2022sdy} likelihood for the TT, TE, and EE spectra at $\ell > 30$, based on the Planck-PR4 \texttt{NPIPE} CMB maps~\cite{Planck:2020olo}. As in the PL18 case, we combine it with the \texttt{simall} likelihood~\cite{Planck:2018vyg} for E-mode polarization and the \texttt{Commander} likelihood~\cite{Planck:2018yye} for temperature anisotropies at $\ell < 30$.
\end{itemize}

\begin{table*}[htpb!]
\begin{center}
\resizebox{0.85\linewidth}{!}{
 \begin{tabular}{ l | c | c | c | c }
  & \textbf{PL18} & \textbf{PL18+lensing} & \textbf{CamSpec} & \textbf{CamSpec+lensing} \\
\hline
\boldmath$\Omega_b h^2$ &  $ 0.02251\pm 0.00016 $&$ 0.02248\pm 0.00015 $&$ 0.02230\pm 0.00015 $&$ 0.02227\pm 0.00014$
\\
\boldmath$\Omega_c h^2$ &  $ 0.1191\pm 0.0014 $&$ 0.1187\pm 0.0014 $&$ 0.1189\pm 0.0012 $&$ 0.1188\pm 0.0012$
\\
\boldmath$100\theta_{MC} $ &  $ 1.04105\pm 0.00032 $&$ 1.04105\pm 0.00031 $&$ 1.04087\pm 0.00026 $&$ 1.04084\pm 0.00026$
\\
\boldmath$\tau$ &  $ 0.0539\pm 0.0081 $&$ 0.0515\pm 0.0082 $&$ 0.0511\pm 0.0083 $&$ 0.0496\pm 0.0081$
\\
\boldmath${\rm{ln}}(10^{10} A_s)$ &  $ 3.041\pm 0.016 $&$ 3.035^{+0.016}_{-0.015} $&$ 3.032\pm 0.017 $&$ 3.029\pm 0.016$
\\
\boldmath$\epsilon$ &  $ 0.0170^{+0.0057}_{-0.0028} $&$ 0.0141^{+0.0074}_{-0.0045} $&$ 0.0159^{+0.0069}_{-0.0035} $&$ 0.0144^{+0.0076}_{-0.0046}$
\\
\boldmath$\delta$ &  $ > 0.0142 $&$ > 0.00650 $&$ > 0.0101 $&$ > 0.00562$
\\
\boldmath$\Omega_\mathcal{K}$ &  $ -0.026^{+0.012}_{-0.011} $&$ -0.0084^{+0.0062}_{-0.0036} $&$ -0.0160^{+0.011}_{-0.0063} $&$ -0.0078^{+0.0060}_{-0.0032}$
\\
\hline
$n_s$ &  $ 0.9671\pm 0.0044 $&$ 0.9680\pm 0.0044 $&$ 0.9656\pm 0.0042 $&$ 0.9656\pm 0.0042$
\\
$\Omega_m$ &  $ 0.418\pm 0.043 $&$ 0.344^{+0.015}_{-0.022} $&$ 0.379^{+0.026}_{-0.043} $&$ 0.344^{+0.014}_{-0.021}$
\\
$H_0$ [km/s/Mpc] &  $ 58.5^{+2.7}_{-3.5} $&$ 64.3^{+2.0}_{-1.5} $&$ 61.4^{+3.2}_{-2.6} $&$ 64.3^{+2.0}_{-1.4}$
\\
$S_8$ &  $ 0.934^{+0.044}_{-0.038} $&$ 0.855^{+0.017}_{-0.020} $&$ 0.891^{+0.031}_{-0.039} $&$ 0.853^{+0.017}_{-0.019}$
\\
\hline
$\Delta \chi^2$ & $-1.73$ & $-1.18$ & $0.49$ & $0.20$
\\
\hline
\end{tabular}
}
\caption{Constraints at 68\%~CL from different Planck likelihood combinations (PL18, PL18+lensing, 
CamSpec, and CamSpec+lensing) on the extended parameter space including 
$\Omega_{\mathcal{K}},\, \epsilon,\, \delta$, together with the usual $\Lambda$CDM parameters and derived 
quantities ($\Omega_m,\, H_0,\, S_8$). The last row reports the difference in the best-fit $\chi^2$, 
defined as $\Delta\chi^2 = \chi^2(\Omega_{\mathcal{K}}) - \chi^2(\epsilon, \delta, \Omega_{\mathcal{K}})$.}
\label{tab:1}
\end{center}
\end{table*}

\begin{table*}[htpb!]
\begin{center}
\resizebox{0.85\linewidth}{!}{
\begin{tabular}{ l | c | c | c | c }
& \textbf{PL18} & \textbf{PL18+lensing} & \textbf{CamSpec} & \textbf{CamSpec+lensing} \\
\hline
\boldmath$\Omega_b h^2$ &  $ 0.02256\pm 0.00016 $&$ 0.02250\pm 0.00016 $&$ 0.02234\pm 0.00015 $&$ 0.02229\pm 0.00014$
\\
\boldmath$\Omega_c h^2$ &  $ 0.1184\pm 0.0014 $&$ 0.1184\pm 0.0014 $&$ 0.1184\pm 0.0012 $&$ 0.1185\pm 0.0012$
\\
\boldmath$100\theta_{MC} $ &  $ 1.04113\pm 0.00032 $&$ 1.04109\pm 0.00031 $&$ 1.04092\pm 0.00026 $&$ 1.04088\pm 0.00027$
\\
\boldmath$\tau$ &  $ 0.0499\pm 0.0078 $&$ 0.0493\pm 0.0080 $&$ 0.0477^{+0.0083}_{-0.0074} $&$ 0.0475^{+0.0084}_{-0.0070}$
\\
\boldmath${\rm{ln}}(10^{10} A_s)$ &  $ 3.032\pm 0.016 $&$ 3.029\pm 0.016 $&$ 3.024^{+0.017}_{-0.015} $&$ 3.024^{+0.017}_{-0.015}$
\\
\boldmath$n_s$ &  $ 0.9695\pm 0.0045 $&$ 0.9692\pm 0.0045 $&$ 0.9676\pm 0.0043 $&$ 0.9667\pm 0.0044$
\\
\boldmath$\Omega_\mathcal{K}$ &  $ < -0.0304 $&$ -0.0107^{+0.0070}_{-0.0047} $&$ -0.024^{+0.013}_{-0.010} $&$ -0.0099^{+0.0071}_{-0.0043}$
\\
\hline
$\Omega_m$ &  $ 0.450^{+0.049}_{-0.028} $&$ 0.352^{+0.018}_{-0.024} $&$ 0.408^{+0.040}_{-0.047} $&$ 0.351^{+0.018}_{-0.024}$
\\
$H_0$ [km/s/Mpc] &  $ 56.3^{+1.3}_{-3.3} $&$ 63.5\pm 2.0 $&$ 59.1^{+2.9}_{-3.5} $&$ 63.6^{+2.2}_{-1.8}$
\\
$S_8$ &  $ 0.956^{+0.040}_{-0.026} $&$ 0.860\pm 0.020 $&$ 0.914\pm 0.038 $&$ 0.858^{+0.018}_{-0.021}$
\\
\hline
\end{tabular}
}
\caption{Constraints at 68\%~CL on $\Omega_{\mathcal{K}}$ from different Planck likelihood combinations 
(PL18, PL18+lensing, CamSpec, and CamSpec+lensing), using the standard \texttt{CAMB} implementation of 
positive curvature. We also report the corresponding derived parameters ($\Omega_m,\, H_0,\, S_8$).}
\label{tab:original}
\end{center}
\end{table*}

In closing this section, we stress that, although $\Omega_{\mathcal{K}}$ can also be constrained by complementary distance measurements such as BAO or SNIa, our main goal here is not to provide the tightest bounds from a combination of different probes. Rather, we aim to quantify to what extent phenomenological or gauge-dependent parametrisations of the primordial spectrum affect the curvature constraints extracted from the CMB. For this reason, in the main text we restrict our analysis to CMB temperature and polarisation anisotropy measurements alone, which makes it easier to isolate the role of both the lensing anomaly and the large-scale anomalies in shaping the inferred value of $\Omega_{\mathcal{K}}$. For completeness, results including low-redshift probes are presented in Appendix~\ref{appendix:A}.

\section{Results}
\label{sec:results} 

\begin{figure}[htpb!]
\centering
\includegraphics[width=0.85\columnwidth]{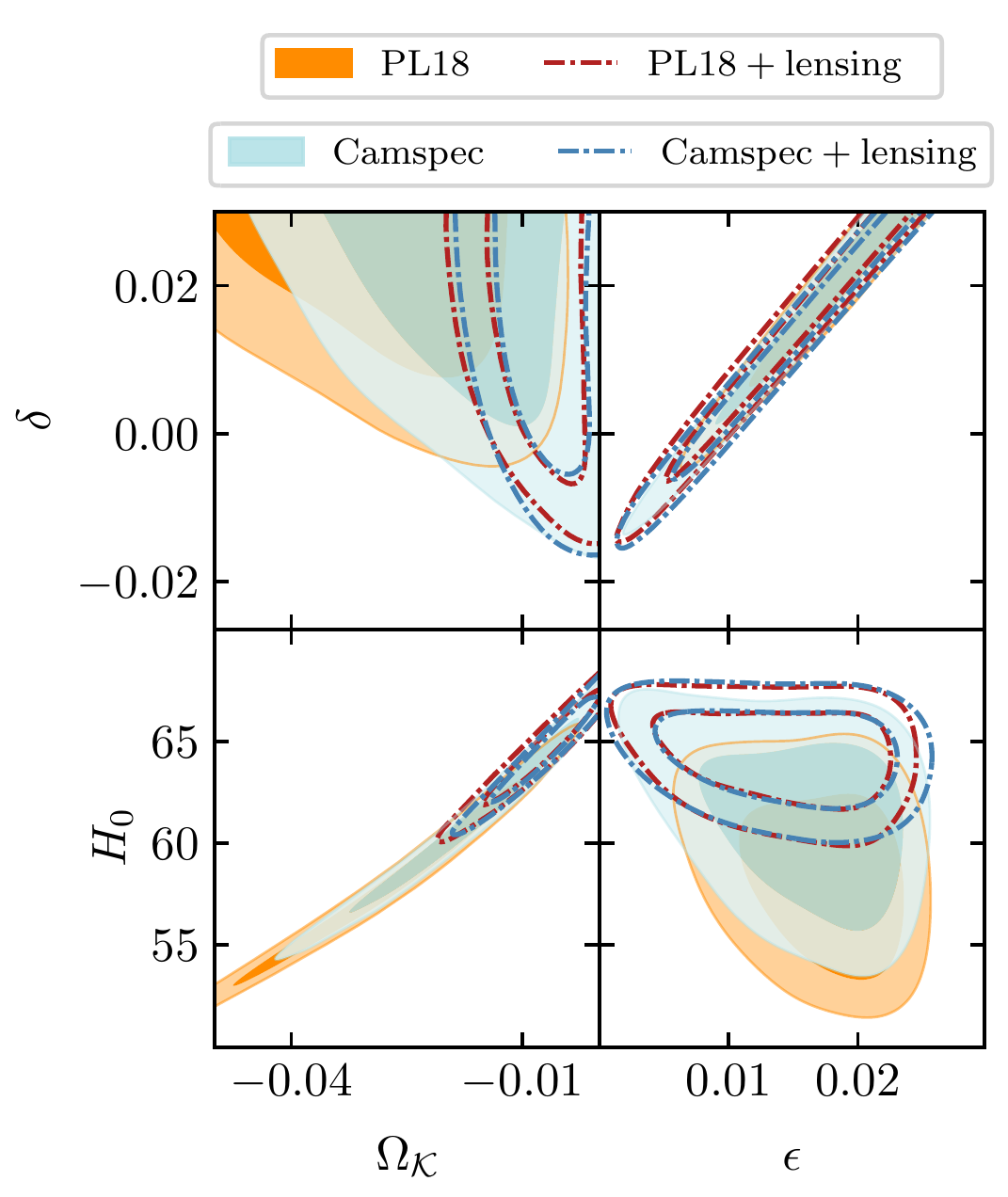}
\caption{Two-dimensional marginalized constraints at 68\% and 95\%~CL on 
$\Omega_{\mathcal{K}},\, \epsilon,\, \delta$ in the gauge-invariant closed-universe model, from different Planck likelihood combinations (PL18, PL18+lensing, CamSpec, and CamSpec+lensing).}
\label{fig:CMB}
\end{figure}

\begin{figure}[htpb!]
\centering
\includegraphics[width=0.85\columnwidth]{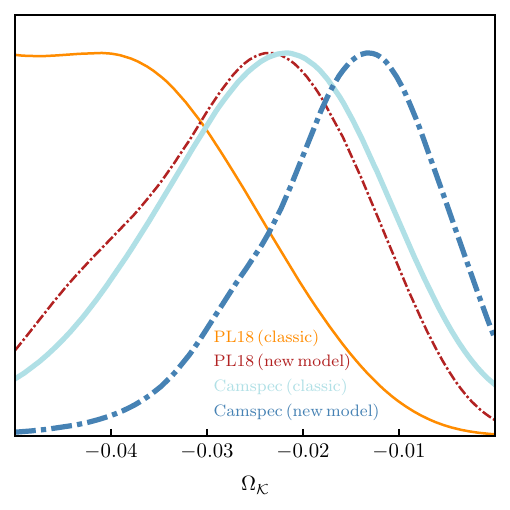}
\caption{One-dimensional marginalized constraints at 68\%~CL on $\Omega_{\mathcal{K}}$ from PL18 and CamSpec likelihoods, comparing the standard \texttt{CAMB} implementation of curvature with our gauge-invariant closed-universe model.}
\label{fig:comparison_1D}
\end{figure}

\begin{figure*}[htpb!]
\centering
\includegraphics[width=0.7\textwidth]{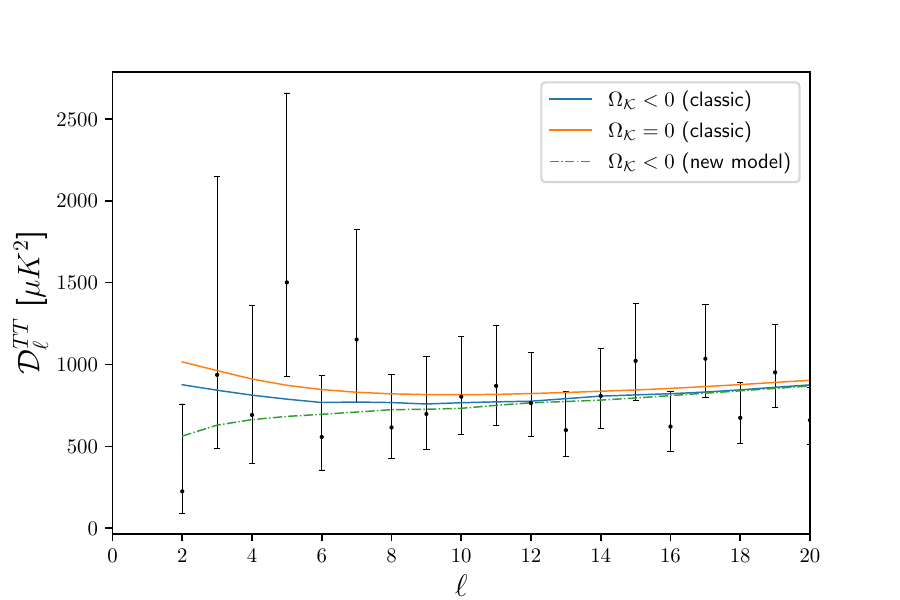}
\caption{Comparison of the best-fit CMB temperature spectra at low multipoles:
$\Lambda$CDM (orange, solid line), the standard curvature implementation
(blue, solid line), and our gauge-invariant positive-curvature model
(green, dash-dotted line). The data points are the Planck~2018 TT spectrum
measurements.}
\label{fig:comparison_data}
\end{figure*}

In this section we present the cosmological constraints obtained with the methodology and datasets described in Sec.~\ref{sec:data}. Unless otherwise stated, all quoted uncertainties correspond to 68\% CL. We work within the extended parameter space of our gauge-invariant closed-universe model, in which the
curvature parameter $\Omega_{\mathcal{K}}$ is sampled together with the slow-roll parameters $\epsilon$ and $\delta$, in addition to the usual $\Lambda$CDM parameters.  

We consider different combinations of Planck likelihoods, namely PL18, PL18+lensing, CamSpec, and CamSpec+lensing, in order to assess
the impact of the lensing anomaly and of alternative high-$\ell$ likelihood choices. The CMB-only constraints on our parameterization are reported in Table~\ref{tab:1}. For comparison, we show in Table~\ref{tab:original} the corresponding results obtained with the standard \texttt{CAMB} implementation of curvature. The two-dimensional marginalized posteriors for $\{\Omega_{\mathcal{K}},\,\epsilon,\,\delta\}$
are displayed in Fig.~\ref{fig:CMB}, while Fig.~\ref{fig:comparison_1D} compares the one-dimensional marginalized constraints on $\Omega_{\mathcal{K}}$ between our implementation and the \texttt{CAMB} curvature treatment. Finally, Fig.~\ref{fig:comparison_data} compares the three best-fit temperature spectra at low multipoles ($\ell \lesssim 20$), corresponding to $\Lambda$CDM, the standard
$\Omega_{\mathcal{K}}$ implementation, and our gauge-invariant parametrization, against the Planck~2018 data. Note that, for completeness, we also perform an extended analysis including geometric measurements from DESI and Pantheon+. The corresponding triangle plot is shown in Fig.~\ref{fig:triangular} and discussed in Appendix A, together with Table~\ref{tab:app} and the complete CMB-only 2D contours in Fig. \ref{fig:triangular_app}.

We begin by discussing the constraints from the PL18 likelihood. In the standard \texttt{CAMB} implementation, the data prefer a more negative value of the curvature parameter, with the posterior reaching the prior boundary and yielding only an
upper limit, $\Omega_{\mathcal{K}} < -0.0304$ at 68\%~CL. This corresponds to a $\sim 3.4\sigma$ indication for a closed universe~\cite{DiValentino:2019qzk,Handley:2019tkm}. In contrast, in our gauge-invariant parametrization we obtain a bounded constraint,
$\Omega_{\mathcal{K}} = -0.026^{+0.012}_{-0.011}$ at 68\%~CL. As shown in Fig.~\ref{fig:comparison_1D}, this result is slightly more consistent with spatial flatness, corresponding to a $\sim 2.5\sigma$ preference for positive curvature.  In our parametrization, the slow-roll parameters are constrained to
$\epsilon = 0.0170^{+0.0057}_{-0.0028}$ and $\delta > 0.0142$ at 68\%~CL while for the other $\Lambda$CDM parameters, the differences between the two approaches are modest. $\tau$ shifts upward by about $0.5\sigma$, in correlation with a higher amplitude $A_s$, while $n_s$ shifts downward by $0.5\sigma$. The remaining free parameters remain stable. As in the standard
analysis, the derived parameters remain in tension with late-time measurements, with $H_0 < 60\,{\rm km\,s^{-1}\,Mpc^{-1}}$, $\Omega_m > 0.4$, and $S_8 > 0.93$.  This behavior can be understood from the strong positive correlation between $\Omega_{\mathcal{K}}$ and $H_0$ visible in Fig.~\ref{fig:CMB}: more negative curvature values imply a smaller Hubble constant. In contrast, $\epsilon$ and $\delta$ are positively correlated with each other but show no significant correlation with either $\Omega_{\mathcal{K}}$ or $H_0$. We quantify the relative goodness of fit by reporting $\Delta\chi^2 \equiv \chi^2(\Omega_{\mathcal{K}}) - \chi^2(\epsilon,\delta,\Omega_{\mathcal{K}})$,
which highlights whether the inclusion of the slow-roll parameters $\epsilon$ and $\delta$ improves the description of the data.
The comparison of the best-fit values shows that the standard \texttt{CAMB} implementation provides a slightly better fit, with $\Delta\chi^2 = -1.73$.  
Nevertheless, as illustrated in Fig.~\ref{fig:comparison_data}, while the standard curvature case already produces a suppression of power at large scales, our parametrization provides a better match to the observed low multipoles, particularly in reproducing the suppressed quadrupole. The fact that our parametrisation improves the fit to the lowest Planck multipoles (better capturing the large-scale power suppression anomaly) while still producing an overall worsening of the global fit implies that it slightly degrades the fit at high-$\ell$ relative to the standard phenomenological parametrisation. The reason is that, in closed quadratic inflation, the exact (gauge-invariant) spectrum ties curvature corrections to the background dynamics, and the posterior favours less negative values of $\Omega_{\mathcal{K}}$ (i.e.\ a geometry closer to flatness) than in the heuristic case. The very negative $\Omega_{\mathcal{K}}$ values that help explain the lensing anomaly at high-$\ell$ are therefore not allowed within our inflationary model, leading to a poorer fit to the high-$\ell$ \texttt{plik} PR3 spectra. Therefore this exercise teaches us an important lesson: when computing the primordial spectrum within physically motivated closed-inflation models, large- and small-scale anomalies in the Planck CMB data do not necessarily pull in the same direction. In our case, power-law inflation reproduces the low-$\ell$ suppression more faithfully, consistent with the theoretical spectrum derived in Ref.~\cite{Kiefer:2021iko}, but at the cost of a reduced ability to accommodate highly negative values of $\Omega_{\mathcal{K}}$ and hence a weaker fit to the lensing anomaly.

When adding the lensing likelihood to PL18, the posterior of the curvature parameter shifts towards zero in both the standard and our gauge-invariant parametrizations, as expected. The resulting constraints are very similar, with $\Omega_{\mathcal{K}} = -0.0107^{+0.0070}_{-0.0047}$ in the standard \texttt{CAMB}
implementation and $\Omega_{\mathcal{K}} = -0.0084^{+0.0062}_{-0.0036}$ in our model. The derived parameters also shift slightly in the same direction as the curvature, but the effect is not sufficient to alleviate the tensions with late-time probes, since $H_0$ remains below the $\Lambda$CDM value while $S_8$ remains higher. In our parametrization, the slow-roll parameter $\epsilon$ is constrained to $\epsilon = 0.0141^{+0.0074}_{-0.0045}$, with a broader posterior than in the PL18-only case. As visible in the two-dimensional posteriors of Fig.~\ref{fig:CMB}, this is mainly a projection effect due to correlations. The parameter $\delta$ remains bounded from below, with $\delta > 0.0065$ at 68\%~CL. The $\Lambda$CDM parameters remain essentially unchanged with respect to the PL18-only case.  The comparison of best fits shows that the original \texttt{CAMB} curvature treatment still provides a slightly better fit, with $\Delta\chi^2 = -1.18$.

We next consider the CamSpec likelihood, which is known to provide a reduced indication for curvature compared to PL18. This behavior is confirmed in both the standard \texttt{CAMB} analysis and in our parametrization. As shown in the one-dimensional posteriors of Fig.~\ref{fig:comparison_1D}, our model shifts the
curvature parameter further towards consistency with flatness, with a preference of about $2\sigma$ for positive curvature. The slow-roll parameters are constrained to $\epsilon = 0.0159^{+0.0069}_{-0.0035}$ and $\delta > 0.0101$ at 68\%~CL while the derived parameters lie between the values obtained with PL18 alone and PL18+lensing, reflecting the intermediate behavior of the curvature constraint in this likelihood. The shifts (or stability) of the other $\Lambda$CDM parameters follow the same trends observed with PL18, with no major deviations. A noticeable difference however is that, in this case, the comparison of best fits slightly favors our parametrization, with $\Delta\chi^2 = +0.49$. While this difference is not statistically significant (especially given that our model includes two additional parameters), it can be understood in light of the behaviour of the CamSpec PR4 spectra, where the lensing anomaly is still present but strongly reduced compared to \texttt{Plik}. This residual preference for enhanced smoothing of the acoustic peaks at high-$\ell$ still translates into a mild preference for a closed universe, but with less negative values of $\Omega_{\mathcal{K}}$ than in Planck PR3. In this regime, the values of $\Omega_{\mathcal{K}}$ allowed by our model not only improve the fit to the lowest multipoles capturing the power suppression anomaly (note that the \texttt{Commander} likelihood encompassing temperature anisotropies at $\ell < 30$ remains common to both combinations) but also provide a satisfactory fit to the higher multipoles. As a result, the improvement at low-$\ell$ is no longer offset by a significant deterioration at high-$\ell$, yielding a small net improvement and reinforcing the interpretation outlined above.

When the lensing likelihood is added to CamSpec, the differences between the two high-$\ell$ likelihoods are effectively washed out. The constraints from CamSpec+lensing and PL18+lensing are nearly identical, with
$\Omega_{\mathcal{K}} = -0.0078^{+0.0060}_{-0.0032}$ in our parametrization, $\epsilon = 0.0144^{+0.0076}_{-0.0046}$, and $\delta > 0.00562$ at 68\%~CL. The comparison of best-fit values yields a slight preference for our model, with $\Delta\chi^2 = +0.20$, though this difference is not statistically significant given the two additional parameters. In fact this can also be regarded as an additional consistency test: once lensing is included, the residual preference for negative curvature in CamSpec is further reduced relative to PL18+lensing, and the constraints inevitably move towards the $\Omega_{\mathcal{K}} \to 0$ limit where all parametrisations become effectively indistinguishable, consistently recovering the flat primordial power spectrum.

Finally, we note once again that the results outlined above are valid for the approximate form of the primordial power spectrum $\mathcal{P}_{\zeta}$ in Eq.(16). However, as explained in Sec.(\ref{sec:param}) and in Appendix B, this approximation holds quite well at the CMB scales considered, and we expect that any impact on the error bars found above should be a negligible second order effect.

\section{Discussion and Conclusion}
\label{sec:conclusion}

In this work, we have revisited the Planck Cosmic Microwave Background constraints on spatial curvature within a quadratic model of closed inflation, where the primordial power spectrum of scalar modes is calculated in a theoretically well-motivated, gauge-invariant framework.

Our work aims to clarify two issues that might appear as technical subtleties but in fact can hold important consequences when inferring constraints on the curvature density parameter. The first is the phenomenological nature of the primordial spectrum for a closed universe usually adopted in standard analyses. The widely used parametrisation~\eqref{eq:spec_efs} can heuristically capture the modifications induced by curvature on the largest scales, where a closed geometry leads to discrete eigenmodes and non-trivial deviations from the flat case. However, while convenient, it does not guarantee that, within concrete inflationary models, the relation between the primordial spectrum and background quantities such as the present-day spatial curvature correctly reproduces the imprints of the inflationary dynamics at those scales. The second is the possible gauge dependence present in many alternative formulations of the primordial spectrum in closed geometries, which can lead to different predictions at large scales.
 
By focusing on quadratic inflation in a closed universe, the primordial spectrum can be calculated in a fully gauge-invariant manner~\cite{Kiefer:2021iko}. We implemented this formulation in \texttt{CAMB}, extended the $\Lambda$CDM parameter space to include the curvature parameter $\Omega_{\mathcal{K}}$ and the two slow-roll parameters $\epsilon$ and $\delta$, and constrained them using Planck CMB data.

We considered different combinations of Planck likelihoods (PL18, PL18+lensing, CamSpec, and CamSpec+lensing) to test the robustness of the apparent preference for positive curvature. Our main numerical findings can be summarized as follows:
\begin{itemize}
    \item Analyzing the dataset combination referred to as PL18 (based on the \texttt{plik} PR3 high-$\ell$ spectra) within our quadratic model of closed inflation, we find a weaker $\sim 2.5\sigma$ preference for a closed universe compared to that obtained using the standard parametrisation~\eqref{eq:spec_efs} of the primordial power spectrum. This is clearly reflected in the shift towards spatial flatness (i.e.\ $\Omega_{\mathcal{K}} \to 0$) observed in the one-dimensional posterior distributions for $\Omega_{\mathcal{K}}$ shown in Fig.~\ref{fig:comparison_1D}. We also note that, for this dataset, our model yields a slightly worse fit compared to the standard parametrisation. This result highlights the importance of calculating and implementing primordial predictions consistently within predictive models of closed inflation, as this can modify both the statistical preference for a closed universe and the final constraints on $\Omega_{\mathcal{K}}$.

    \item Analyzing the dataset combination referred to as CamSpec (based on the \texttt{CamSpec} PR4 high-$\ell$ spectra) within our quadratic model of closed inflation, the preference for positive curvature is further reduced, to about $2\sigma$, while the slow-roll parameters remain constrained at meaningful levels. This outcome is somewhat expected, since the indication for curvature is weaker in \texttt{CamSpec} than in \texttt{plik}. For this combination, our model also yields a slight improvement in the fitting of the Planck data compared to the standard parametrisation (although the difference is not statistically significant given the larger number of parameters).
    
    \item Adding Planck lensing data (to both PL18 and CamSpec), we observe a further shift of the posteriors in both models towards $\Omega_{\mathcal{K}} = 0$, as expected. In this regime, the two parametrisations become broadly consistent, since both primordial spectra converge to the standard flat case.
\end{itemize}

Beyond the numerical constraints, for both PL18 and CamSpec our parametrisation modifies the shape of the CMB temperature spectrum at large angular scales, yielding an improved description of the low-$\ell$ power suppression anomaly observed at multipoles $\ell < 10$, see Fig.~\ref{fig:comparison_data}. In particular, the low quadrupole is better reproduced compared to the standard \texttt{CAMB} curvature implementation, highlighting the physical impact of our gauge-invariant treatment of the primordial spectrum. However, as already noted, this improvement at large scales does not necessarily translate into an overall better fit compared to the standard closed parametrisation. While in CamSpec we observe a small improvement in the global fit, in PL18 we find a slight worsening.  This behaviour can be traced back to the different constraints on spatial curvature. In closed quadratic inflation, the exact (gauge-invariant) spectrum ties curvature corrections to the background dynamics, and the posterior favours less negative values of $\Omega_{\mathcal{K}}$ (i.e.\ a geometry closer to flatness) than in the heuristic case. The very negative values of $\Omega_{\mathcal{K}}$ that help explain the lensing anomaly at high-$\ell$ are therefore not allowed within our inflationary model, leading to a poorer fit to the high-$\ell$ \texttt{plik} PR3 spectra. On the other hand, in the \texttt{CamSpec} high-$\ell$ PR4 spectra the preference for a smoothing excess of the acoustic peaks is reduced, and the less negative values of $\Omega_{\mathcal{K}}$ favored by our closed-inflation model are sufficient to yield an overall improvement.

As for the other parameters, we observe minimal to non-existent shift in their best-fit values, and all the derived parameters ($H_0$, $\Omega_m$, $S_8$) remain in tension with late-time probes.

Overall, our case study demonstrates that deriving the primordial spectrum consistently within closed-inflation models -- where the calculation is fully gauge-invariant and quantities such as the primordial spectrum, inflationary parameters, and geometric degrees of freedom are treated on equal footing -- can lead to meaningful differences in the inferred constraints on $\Omega_{\mathcal{K}}$, with implications that extend well beyond the interpretation of the (apparent or spurious) Planck preference for a closed universe. Our broader message is that implementing physically motivated closed-universe models can substantially modify the results that would otherwise be inferred from purely phenomenological parametrisations of the primordial power spectrum.

There also remains considerable scope for theoretical progress. In this work we focused on quadratic inflation, a model that in the flat case is essentially excluded because of its predictions for the tensor-to-scalar ratio, which conflict with current BICEP/Keck limits~\cite{Planck:2018jri,BICEP:2021xfz,Forconi:2021que}. That said, precise calculations for the tensor spectrum in a closed universe have not been derived for this model, preventing us from assessing the impact of spatial curvature on primordial gravitational waves. Building on existing discussions~\cite{Franco:2017pxt,Bonga:2016cje,DAgostino:2023tgm} extending the gauge-invariant treatment to tensor perturbations therefore represents a natural next step. More broadly, applying this framework to other benchmark inflationary scenarios that are successful in the flat case, such as Starobinsky inflation, will be essential for building a more comprehensive picture. These theoretical advances, together with the improved precision of forthcoming Stage-IV CMB experiments, will be crucial to establish with confidence the role of curvature in the cosmological model.


\begin{acknowledgments}
\noindent W.G. is supported by the Lancaster–Sheffield Consortium for Fundamental Physics under STFC grant: ST/X000621/1. E.D.V. is supported by a Royal Society Dorothy Hodgkin Research Fellowship. This article is based upon work from the COST Action CA21136 - ``Addressing observational tensions in cosmology with systematics and fundamental physics (CosmoVerse)'', supported by COST - ``European Cooperation in Science and Technology''.
We acknowledge the IT Services at The University of Sheffield for the provision of services for High Performance Computing.
\end{acknowledgments}

\newpage
\twocolumngrid
\bibliography{refs}

\appendix

\section{Complete model constraints, including low-redshift data}
\label{appendix:A}

\begin{figure*}[htpb!]
\centering
\includegraphics[width=0.7\textwidth]{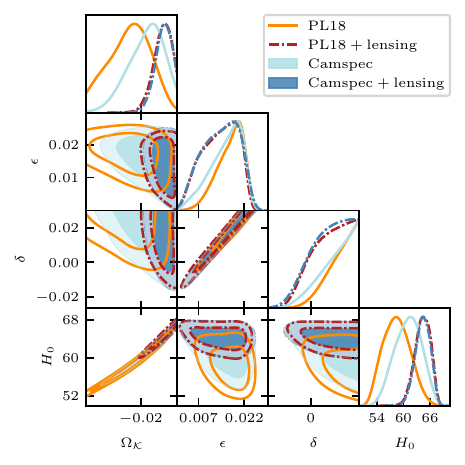}
\caption{One- and two-dimensional marginalized constraints at 68\% and 95\% CL on $H_0$, $\Omega_K$, $\epsilon$, and $\delta$ in the gauge-invariant closed-universe model.}
\label{fig:triangular_app}
\end{figure*}

\begin{table*}[htpb!]
\begin{center}
\resizebox{0.85\linewidth}{!}{

\begin{tabular}{ l | c | c | c | c }

  & \textbf{PL18+lensing} & \textbf{PL18+lensing} & \textbf{CamSpec+lensing} & \textbf{CamSpec+lensing} \\
    &  & \textbf{+DESI+P+} &  & \textbf{+DESI+P+} \\
\hline

\boldmath$\Omega_b h^2$ &  $ 0.02248\pm 0.00015 $&$ 0.02257\pm 0.00013 $&$ 0.02227\pm 0.00014 $&$ 0.02234\pm 0.00012$
\\

\boldmath$\Omega_c h^2$ &  $ 0.1187\pm 0.0014 $&$ 0.11744^{+0.00075}_{-0.00068} $&$ 0.1188\pm 0.0012 $&$ 0.11743^{+0.00072}_{-0.00064}$
\\

\boldmath$100\theta_{MC} $ &  $ 1.04105\pm 0.00031 $&$ 1.04126\pm 0.00028 $&$ 1.04084\pm 0.00026 $&$ 1.04104\pm 0.00023$
\\

\boldmath$\tau$ &  $ 0.0515\pm 0.0082 $&$ 0.0612^{+0.0066}_{-0.0078} $&$ 0.0496\pm 0.0081 $&$ 0.0595^{+0.0064}_{-0.0078}$
\\

\boldmath${\rm{ln}}(10^{10} A_s)$ &  $ 3.035^{+0.016}_{-0.015} $&$ 3.053^{+0.013}_{-0.015} $&$ 3.029\pm 0.016 $&$ 3.048^{+0.013}_{-0.015}$
\\

\boldmath$\epsilon$ &  $ 0.0141^{+0.0074}_{-0.0045} $&$ < 0.0145 $&$ 0.0144^{+0.0076}_{-0.0046} $&$ 0.0111^{+0.0070}_{-0.011}$
\\

\boldmath$\delta$ &  $ > 0.00650 $&$ > -0.00172 $&$ > 0.00562 $&$ 0.007^{+0.015}_{-0.019}$
\\

\boldmath$\Omega_\mathcal{K}$ &  $ -0.0084^{+0.0062}_{-0.0036} $&$ > -0.000583 $&$ -0.0078^{+0.0060}_{-0.0032} $&$ > -0.000511$
\\
\hline
$n_s$ &  $ 0.9680\pm 0.0044 $&$ 0.9713\pm 0.0034 $&$ 0.9656\pm 0.0042 $&$ 0.9691\pm 0.0034$
\\
$\Omega_m$ &  $ 0.344^{+0.015}_{-0.022} $&$ 0.3017\pm 0.0036 $&$ 0.344^{+0.014}_{-0.021} $&$ 0.3032\pm 0.0035$
\\

$H_0$ [km/s/Mpc] &  $ 64.3^{+2.0}_{-1.5} $&$ 68.28\pm 0.28 $&$ 64.3^{+2.0}_{-1.4} $&$ 68.06\pm 0.27$
\\

$S_8$ &  $ 0.855^{+0.017}_{-0.020} $&$ 0.8096\pm 0.0083 $&$ 0.853^{+0.017}_{-0.019} $&$ 0.8094\pm 0.0081$
\\

\hline

\end{tabular}
}
\label{tab:app}
\caption{Constraints at 68\%~CL from different Planck likelihood combinations with the inclusion of 
geometric measurements (DESI and Pantheon+) on the extended parameter space including 
$\Omega_{\mathcal{K}},\, \epsilon,\, \delta$, together with the usual $\Lambda$CDM parameters and derived 
quantities ($\Omega_m,\, H_0,\, S_8$).}
\end{center}
\end{table*}

\begin{figure*}[htpb!]
\centering
\includegraphics[width=0.55\textwidth]{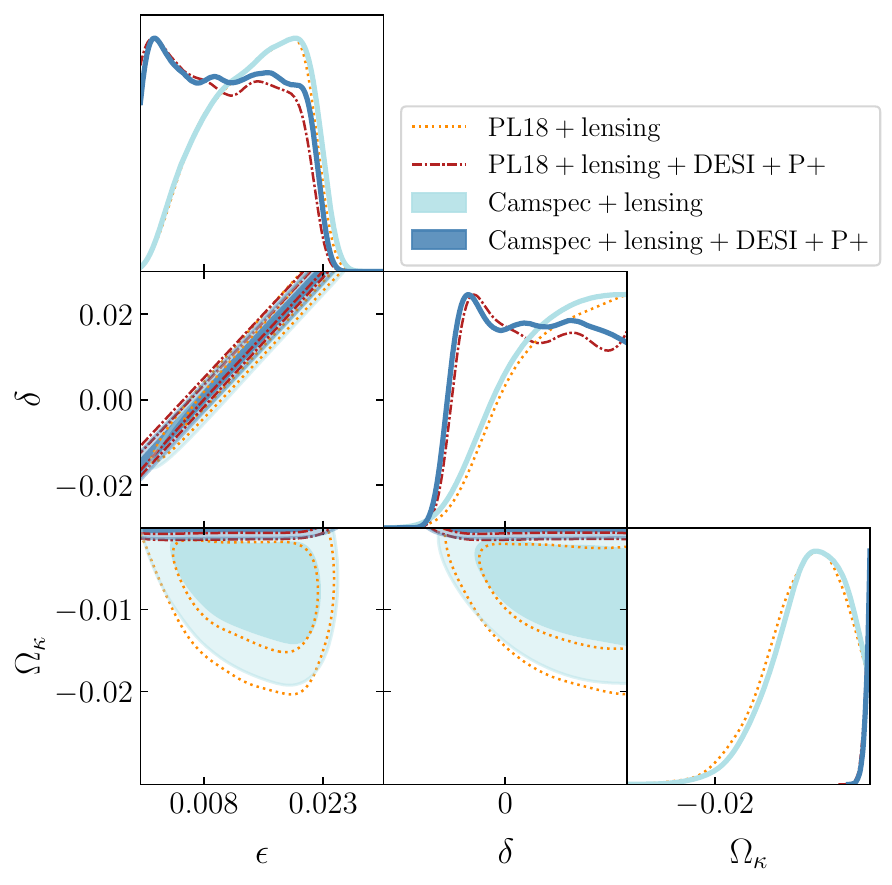}
\caption{One- and two-dimensional marginalized constraints at 68\% and 95\%~CL on $\Omega_{\mathcal{K}},\, \epsilon,\, \delta$ in the gauge-invariant closed-universe model, including geometric measurements from DESI and Pantheon+, combined with Planck likelihoods.}   
\label{fig:triangular}
\end{figure*}

We investigate our parametrization when combining Planck CMB data with low-redshift geometric measurements from DESI and Pantheon+. The inclusion of late-time distance indicators removes the preference for a closed
universe, and in this regime our model becomes not optimal to explain observations since it forces $\Omega_{\mathcal{K}} < 0$. Nevertheless, it is instructive to verify the consistency of the results. Using PL18+lensing together with DESI and Pantheon+, we find results fully consistent
with flat $\Lambda$CDM, both for the primary and derived parameters. In particular, as one can see from Table \ref{tab:app}, we obtain $\Omega_{\mathcal{K}} > -0.000583$, $\epsilon < 0.0145$, and $\delta > -0.00172$ at 68\%~CL. Replacing PL18 with CamSpec leads to very similar constraints, $\Omega_{\mathcal{K}} > -0.000511$, $\epsilon = 0.0111 \pm 0.0034$, and $\delta = 0.007^{+0.015}_{-0.019}$. The corresponding triangular plots are displayed in Fig.~\ref{fig:triangular}, showing that in these cases, fully consistent with a flat universe, the slow-roll parameters $\epsilon$ and $\delta$ are largely unconstrained.

\section{On the validity of the approximation in Eq.(\ref{eq:simply})}
\label{appendix:B}

\begin{figure*}[htpb!]
    \centering
    \includegraphics[width=\linewidth]{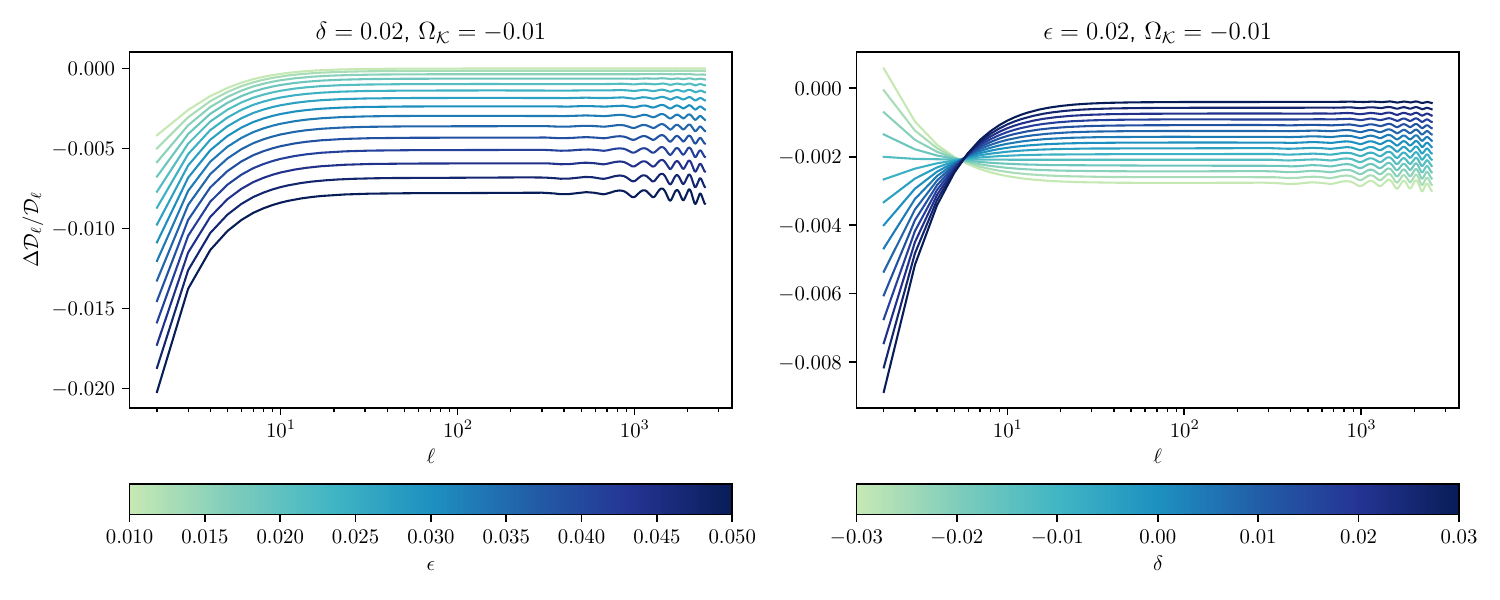}    \caption{Fractional difference between the approximated and non-approximated theoretical predictions of the CMB TT power spectrum for our model. The difference becomes most pronounced as $\epsilon$ increases, but its magnitude does not exceed $\sim 2\%$ for multipoles below $\ell \approx 10$ (in the cosmic-variance-limited regime), and remains below $\sim 1\%$ for multipoles above $\ell \approx 10$.}
    \label{fig:residual}
    \vspace{1cm}
    \centering
    \includegraphics[width=\linewidth]{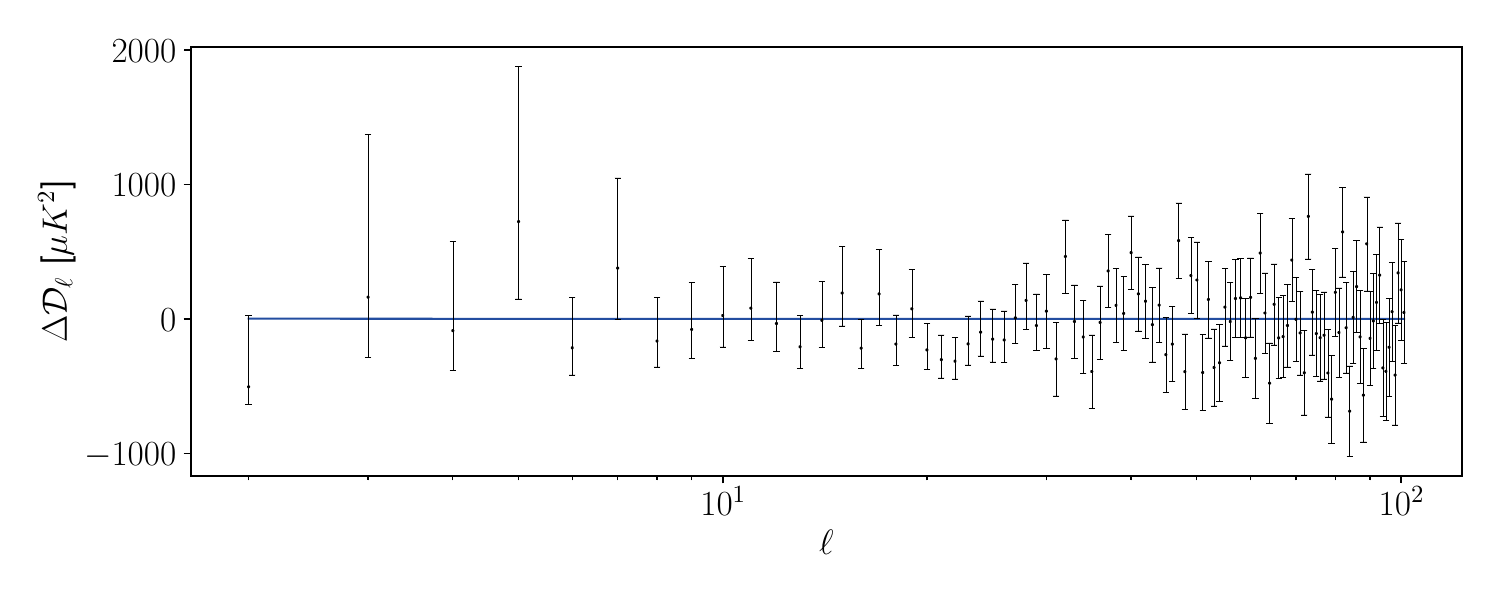}    \caption{Difference between the theoretical predictions of the approximated and non-approximated forms of the CMB TT power spectrum for our model (the nearly horizontal blue line), compared with the Planck 2018 temperature data (points with error bars)~\cite{Planck:2018vyg}. Both versions of the spectrum are computed using the best-fit parameter values obtained for the approximate case with the Planck 2018 data. A difference between the two implementations is present, but it is not large enough to exceed the experimental uncertainties or to induce any tangible systematic offset in the inferred constraints. In particular, the difference between the two cases is almost zero because the best-fit values for $\Omega_{\mathcal{K}}$ of both are quite close to zero, the limit at which the two parametrisations become identical (i.e., both reduce to the spatially flat spectrum of Eq.(\ref{eq:power_law})).}
    \label{fig:residual_data}
\end{figure*}

To validate the robustness of the approximation introduced by Eq.~(\ref{eq:simply}), we have implemented both the approximated (i.e., Eq.~(\ref{fsp1})) and the full (i.e., Eq.~(\ref{eq:spec_full})) expressions of $\mathcal{P}_{\zeta}$ in \texttt{CAMB}, and explicitly compared their impact on the computation of the CMB temperature angular power spectra $\mathcal{D}_\ell$. The results are shown in Fig.~\ref{fig:residual}. For values of $\epsilon$ of order $\epsilon \sim 0.01$ and $\delta \sim 0.02$ (i.e., of the order of the best-fit values quoted in the manuscript), the relative difference between the approximate and full implementations is less than $0.5\%$ on the largest scales, and becomes negligible at smaller scales. Altering the value of $\delta$ has only a minimal impact, see the right panel. Similarly, when increasing $\epsilon$ to values up to three times larger than those preferred by the MCMC analysis, the aforementioned difference between the two implementations remains below $0.5\%$ at small scales and approaches $1.5\%$ at large scales. However, given the tight uncertainties on $\epsilon$ reported in the manuscript, such values already lie well outside the region favoured by the analysis. Only for values of $\epsilon$ as large as five times the best-fit region does the discrepancy reach the percent level, still remaining $\sim 1\%$ at small scales. In this extreme regime, the largest deviations, reaching at most $\sim 2\%$, occur at very low multipoles. However, on the one hand, these correspond to scales where Planck temperature data are dominated by cosmic variance and therefore have comparatively large error bars, as illustrated in Fig.~\ref{fig:residual_data}. On the other hand, these values of $\epsilon$ are strongly excluded by the data and are therefore never probed in our MCMC runs. As a result, across the entire parameter range explored by the present analysis, the uncertainty induced by our approximation is always negligible compared to the observational uncertainties and cannot lead to any appreciable effect on the final constraints. Additionally, the region where our approximations more faithfully reproduce the non-approximated spectrum is found to be exactly coincident with the region of parameter space favoured by the data. This is seen in Fig.~\ref{fig:residual_data}: when considering the best-fit parameter values (i.e., in the region of maximum likelihood), the difference between the approximate and full implementations is completely indistinguishable. We therefore conclude that, since the implementation differences are well within experimental errors, it is safe to neglect $\lambda_{\epsilon,\delta}(n)$ and the terms proportional to $\epsilon$ in $f_{\epsilon}(n)$.

 \end{document}